\documentclass[12pt]{article} 
\usepackage[utf8x]{inputenc} 
\usepackage[T1]{fontenc} 
\usepackage{amsmath,amssymb,bm,wasysym,xfrac} 
\usepackage{xcolor} 
\usepackage{graphicx} 
\usepackage{cite} 

\usepackage{sectsty} 
\usepackage{geometry} 
\usepackage{fancyhdr} 
\usepackage{setspace} 
\usepackage[nottoc,notlof,notlot]{tocbibind} 

\usepackage[unicode]{hyperref} 
\hypersetup{bookmarksnumbered=true, bookmarksopen=true, bookmarksopenlevel=2, breaklinks=true, citecolor=blue, colorlinks=true, linkcolor=red, linktoc=page, pdfborder={0 0 0}, pdfcreator={}, pdfproducer={}, pdfstartview=FitH, plainpages=false, unicode=true, urlcolor=blue}

\DeclareUnicodeCharacter{"0393}{\Gamma}
\DeclareUnicodeCharacter{"0394}{\Delta}
\DeclareUnicodeCharacter{"0398}{\Theta}
\DeclareUnicodeCharacter{"039B}{\Lambda}
\DeclareUnicodeCharacter{"039E}{\Xi}
\DeclareUnicodeCharacter{"03A0}{\Pi}
\DeclareUnicodeCharacter{"03A3}{\Sigma}
\DeclareUnicodeCharacter{"03A5}{\Upsilon}
\DeclareUnicodeCharacter{"03A6}{\Phi}
\DeclareUnicodeCharacter{"03A8}{\Psi}
\DeclareUnicodeCharacter{"03A9}{\Omega}
\DeclareUnicodeCharacter{"03B1}{\alpha}
\DeclareUnicodeCharacter{"03B2}{\beta}
\DeclareUnicodeCharacter{"03B3}{\gamma}
\DeclareUnicodeCharacter{"03B4}{\delta}
\DeclareUnicodeCharacter{"03B5}{\epsilon}
\DeclareUnicodeCharacter{"03B6}{\zeta}
\DeclareUnicodeCharacter{"03B7}{\eta}
\DeclareUnicodeCharacter{"03B8}{\theta}
\DeclareUnicodeCharacter{"03D1}{\vartheta}
\DeclareUnicodeCharacter{"03B9}{\iota}
\DeclareUnicodeCharacter{"03BA}{\kappa}
\DeclareUnicodeCharacter{"03BB}{\lambda}
\DeclareUnicodeCharacter{"03BC}{\mu}
\DeclareUnicodeCharacter{"03BD}{\nu}
\DeclareUnicodeCharacter{"03BE}{\xi}
\DeclareUnicodeCharacter{"03C0}{\pi}
\DeclareUnicodeCharacter{"03C1}{\rho}
\DeclareUnicodeCharacter{"03C3}{\sigma} 
\DeclareUnicodeCharacter{"03C4}{\tau}
\DeclareUnicodeCharacter{"03C5}{\upsilon}
\DeclareUnicodeCharacter{"03C6}{\phi}
\DeclareUnicodeCharacter{"03D5}{\varphi}
\DeclareUnicodeCharacter{"03C7}{\chi}
\DeclareUnicodeCharacter{"03C8}{\psi}
\DeclareUnicodeCharacter{"03C9}{\omega}
\DeclareUnicodeCharacter{"21D0}{\Leftarrow}
\DeclareUnicodeCharacter{"0212B}{\AA}
\DeclareUnicodeCharacter{"00B7}{\cdot}
\DeclareUnicodeCharacter{"266A}{\eighthnote}
\DeclareUnicodeCharacter{"266B}{\twonotes}
\PrerenderUnicode{ä}\PrerenderUnicode{×}

\definecolor{title}{rgb}{0.1,0.5,0.9}
\definecolor{abst}{rgb}{0.366,0.366,0.366}
\definecolor{sect}{rgb}{0.2,0.4,0.8}
\definecolor{ssect}{rgb}{0.3,0.3,0.7}
\definecolor{sssect}{rgb}{0.4,0.2,0.6}
\definecolor{appsect}{rgb}{0.5,0.1,0.5}
\definecolor{ref}{rgb}{0.0,0.0,1.0}
\definecolor{orcidlogocol}{HTML}{A6CE39}
\newcommand{\Title}[1] {\title{\color{title}\Huge #1}}
\newcommand{\TPheader}[3] {\date{}\maketitle\thispagestyle{fancy}\pagenumbering{alph}\lhead{#1}\chead{#2}\rhead{#3}\cfoot{}}

\newcommand{\Abstract}[1] {\begin{abstract}\normalsize #1 \end{abstract}}
\newcommand{\makepage}[1] {\newpage\pagenumbering{#1}}

\sectionfont{\color{sect}}
\subsectionfont{\color{ssect}}
\subsubsectionfont{\color{sssect}}
\renewcommand{\appendix}{\setcounter{section}{0}\sectionfont{\color{appsect}}\renewcommand{\thesection}{\Alph{section}}\renewcommand*{\theHsection}{app.\the\value{section}}} 
\newcommand\references[1]{\sectionfont{\color{ref}}\bibliographystyle{hephys}\bibliography{#1}}

\newcommand\eqs[1] {\begin{align}#1\end{align}}
\newcommand\eqsn[1] {\begin{align*}#1\end{align*}}
\newcommand\eqss[1] {\begin{align}\begin{split}#1\end{split}\end{align}}
\newcommand\eqst[1] {\begin{multline}#1\end{multline}}

\newcommand\eqsc[1] {\begin{gather}#1\end{gather}}

\newcommand\eqsg[1] {\equ{\begin{aligned}#1\end{aligned}}}
\newcommand\equ[1] {\begin{equation}#1\end{equation}}
\newcommand\equn[1] {\begin{equation*}#1\end{equation*}}
\newcommand\fig[2] {\begin{figure}[#1]\centering #2\end{figure}}
\newcommand\pmat[1] {\begin{pmatrix}#1\end{pmatrix}}

\newcommand\half {\tfrac{1}{2}}
\renewcommand\i {\dot{\iota}}

\renewcommand\( {\left(}
\renewcommand\) {\right)}
\newcommand\wh {\widehat}
\DeclareMathOperator{\tr}{tr}

\newcommand\A {{\mathcal A}}

\newcommand\D {{\mathcal D}}
\newcommand\F {{\mathcal F}}

\newcommand\N {{\mathcal N}} 

\renewcommand\S {{\mathcal S}}
\renewcommand\U {{\mathcal U}}

\renewcommand\a {{\mathfrak a}}

\newcommand\Uv[2]{U^{#1}_{~#2}}
\newcommand\Ub[2]{\bar{U}_{#1}^{~#2}}
\newcommand\bk[1] {\langle #1 \rangle}
\newcommand\tint {{\textstyle ∫}}
\newcommand\ie {i.e.}
\newcommand\nn {\nonumber\\}

\numberwithin{equation}{section} 
\begin{document}
\Title{Simplifying 4d $\N=3$ Harmonic Superspace}

\author{\href{mailto:dharmesh.jain@bose.res.in}{Dharmesh Jain}\,\href{https://orcid.org/0000-0002-9310-7012}{\includegraphics[scale=0.0775]{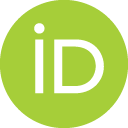}}\\
\emph{\normalsize Department of Theoretical Sciences, S. N. Bose National Centre for Basic Sciences,}\\
\emph{\normalsize Block--JD, Sector--III, Salt Lake City, Kolkata 700106, India}\medskip\\
\href{mailto:cju@nchu.edu.tw}{Chia-Yi Ju}\,\href{https://orcid.org/0000-0001-7038-3375}{\includegraphics[scale=0.0775]{ORCIDiD_icon128x128.png}}\\
\emph{\normalsize Department of Physics, National Chung Hsing University,}\\
\emph{\normalsize 145 Xingda Road, South District, Taichung City 40227, Taiwan}\smallskip\\
and\smallskip\\
\href{mailto:siegel@insti.physics.sunysb.edu}{Warren Siegel}\,\href{http://insti.physics.sunysb.edu/\~siegel/plan.html}{\includegraphics[scale=0.1525]{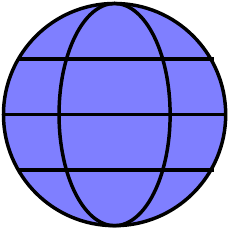}}\\
\emph{\normalsize C. N. Yang Institute for Theoretical Physics, Stony Brook University,}\\
\emph{\normalsize 100 Nicolls Road, Stony Brook, NY 11794-3840, USA}\bigskip\\
}

\TPheader{\today}{}{YITP--SB--2020--22} 

\Abstract{We quantize super Yang-Mills action in $\N=3$ harmonic superspace using ``Fermi-Feynman'' gauge and also develop the background field formalism. This leads to simpler propagators and Feynman rules that are useful in performing explicit calculations. The superspace rules are used to show that divergences do not appear at 1-loop and beyond. We also compute a finite contribution to the effective action from a 4-point diagram at 1-loop, which matches the expected covariant result.
}

\makepage{arabic} 
\tableofcontents

\section{Introduction}
The $\N=3$ harmonic superspace in four dimensions was developed by GIKOS around three and a half decades ago \cite{Galperin:1984bu,Galperin:2007wpa}, and it provided the first successful off-shell formulation of 4d $\N=3$ super Yang-Mills (SYM) theory. This theory was quantized in ``Landau'' gauge a few years later by Delduc and McCabe \cite{Delduc:1988cp}; however, the propagators obtained did not lend themselves to easier calculations. It is well known that the field content of a $\N=3$ vector multiplet is the same as that of a $\N=4$ one, and Zupnik explicitly showed this hidden supersymmetry of the $\N=3$ SYM in \cite{Zupnik:2003us}. The $\N=3$ superspace also manifests the full superconformal symmetry and using such symmetry arguments, low-energy effective action for $\N=3$ and $\N=4$ were considered by Zupnik and collaborators in \cite{Ivanov:2001ec,Buchbinder:2004rj,Buchbinder:2011zu,Buchbinder:2016xeq}. A ``twistorial'' perspective on the $\N=3$ SYM action was presented in \cite{Schwab:2013hf} a few years ago, but no concrete progress has yet been made ``to bring the quantization scheme into a form suitable for computations''\cite{Galperin:2007wpa}.

We present evidence of some progress in the direction of simplifying computations here. We choose ``Fermi-Feynman'' gauge to drastically reduce the number ($9→1$) and simplify the form (\{chiral, antichiral, linear\}-analytic → just analytic) of propagators when compared to \cite{Delduc:1988cp}. This simplifies the proof of the nonrenormalization theorem as one might expect. Moreover, we also introduce the background field formalism in the $\N=3$ harmonic superspace to simplify computations further.

In Section \ref{sec:review}, we review the basic ingredients of the $\N=3$ harmonic superspace and the $\N=3$ SYM action. In Section \ref{sec:FFgauge} we introduce the ``Fermi-Feynman'' gauge to gauge fix this SYM action and derive the propagators. As an application, we prove the nonrenormalization theorem. In Section \ref{sec:BFgauge} we introduce the background field gauge to simplify the diagrammatic computations and present a sample calculation. Finally, we conclude with some discussion in Section \ref{sec:discuss}.

\section{Review}\label{sec:review}
Our notation will closely follow \cite{Delduc:1988cp} and we review it here for orientation purposes. The full 4d $\N=3$ superspace has the usual set of ordinary bosonic ($x^{α\dot{α}}$) and fermionic ($θ_i^α, \bar{θ}^{i\dot{α}}$) coordinates with $i=1,2,3$. The harmonic superspace augments these by six internal bosonic coordinates of the R-symmetry coset $SU(3)/U(1)×U(1)$, denoted collectively as $u$. Using these internal coordinates, an ``analytic'' subspace with eight out of the twelve $θ$'s of the full superspace is identified, which allows one to construct an off-shell action for the $\N=3$ SYM and proceed with its quantization. In this section, we discuss the internal coordinates in some detail first, then the fermionic ones, and finally the superspace action of $\N=3$ SYM. The quantization is dealt with in subsequent sections.

\subsection{Internal Coordinates}
A $SU(3)/U(1)×U(1)$ coset element can be parameterized in matrix form as follows
\eqsc{\U=\pmat{u_i^1,u_i^2,u_i^3}≡\pmat{u_i^{(1,1)},u_i^{(-1,1)},u_i^{(0,-2)}}; \qquad \U^{†}=\pmat{\bar{u}^i_1 \\ \bar{u}^i_2 \\ \bar{u}^i_3}≡\pmat{\bar{u}^{i(-1,-1)} \\ \bar{u}^{i(1,-1)} \\ \bar{u}^{i(0,2)}} \label{ucharges}\\
\begin{gathered}
\text{Constraints: }\; \U^{†}\U=\U\U^{†}=1\,; \qquad \det\U=1 \\
⇒ \bar{u}_a·u^b = \bar{u}^i_au^b_i = δ^b_a\,, \quad u^a_i\bar{u}_a^j = δ^j_i\,; \quad ε^{ijk}u^1_iu^2_ju^3_k=1\,.
\end{gathered} \label{uconstrnts}
}
The notation $(q_1,q_2)$ denotes the charges corresponding to the two Cartan $U(1)$ generators $Q_1,Q_2$ of $SU(3)$. Given the constraints in \eqref{uconstrnts}, we have eight independent coordinates in the $\U$-matrix, as expected for $SU(3)$. However, we also require the two $U(1)$ charges to be fixed (\ie, $Q_i u=q_i u$), which effectively implements the $U(1)^2$ quotient, and we are left with six independent coordinates.

Furthermore, the six harmonic covariant derivatives acting on these coordinates are
\equ{D^a_b = u^a_i\frac{\partial}{\partial u^b_i} - \bar{u}^i_b\frac{\partial}{\partial \bar{u}^i_a} \qquad (a≠b) \,.
}
These derivatives satisfy the $SU(3)$ Lie algebra given by
\equ{[D^a_b,D^c_d] = \delta^c_bD^a_d - \delta^a_dD^c_b\,.
}
We note here that the two Cartan generators are given in terms of $D^a_a$ (no sum over $a$) as follows
\equ{\begin{aligned}
Q_1 &= D^1_1 - D^2_2\\
Q_2 &= D^1_1 + D^2_2 - 2D^3_3
\end{aligned}\qquad \Rightarrow \qquad \begin{aligned}
D^1_1 - D^3_3 &= \tfrac{1}{2}\(Q_1 + Q_2\)\\
D^2_2 - D^3_3 &= \tfrac{1}{2}\(Q_2 - Q_1\).
\end{aligned}
}
Their commutators with the harmonic derivatives are $[Q_i, D^a_b]=q_i D^a_b$ with the charges given by
\equn{\begin{tabular}{c|cccccc}
$U(1)^2$ & $D_3^1$ & $D_2^3$ & $D_2^1$ & $D_1^3$ & $D_3^2$ & $D_1^2$ \\
\hline
$q_1$ & $1$ & $1$ & $2$ & $-1$ & $-1$ & $-2$ \\
$q_2$ & $3$ & $-3$ & $0$ & $-3$ & $3$ & $0$
\end{tabular}
}

In what follows, we will mostly be dealing with functions defined at two different points in this coset space, labelled as $u$ and $v$. We denote their products by the notation $U^a{}_b=u^a·\bar{v}_b$ and $\bar{U}_a{}^b=\bar{u}_a·v^b$ such that the covariant derivatives in this basis simply read
\equ{D^a_b = {U^a}_{c}\frac{\partial}{\partial {U^b}_{c}} - \bar{U}_b{}^c\frac{\partial}{\partial \bar{U}_a{}^c} \qquad (a≠b) \,.
}

Finally, the integration over this coset space is defined such that only a $SU(3)$ singlet integrand gives a non-vanishing result, \ie,
\equ{∫du\,1=1\,;\qquad ∫du\,D^a_bf(u)=0\,.
}
The latter integral allows one to integrate by parts in the $u$-space.

\subsection{Fermionic Coordinates}
We make a coordinate transformation of the usual $θ$'s with $SU(3)$ indices to $θ$'s having definite $U(1)$ charges as follows:
\equ{θ_a^α=\bar{u}^i_aθ^α_i\,,\quad \bar{θ}^{a\dot{α}}=u_i^a\bar{θ}^{i\dot{α}}\,.
}
The index $a$ identifies the $U(1)$ charges straightforwardly via \eqref{ucharges}. Then, the corresponding spinorial covariant derivatives satisfy the following commutators:
\eqsc{\{D^a_α, D_β^b\}=0\,,\quad \{\bar{D}_{a\dot{α}},\bar{D}_{b\dot{β}}\}=0\,,\quad \{D_α^a, \bar{D}_{b\dot{β}}\}=\i δ^a_b ∂_{α\dot{β}}\,, \label{intgstr}\\
[D^a_b,D^c_α]=δ^c_bD^a_α\,,\quad [D^a_b,\bar{D}_{c\dot{α}}]=-δ^a_c\bar{D}_{b\dot{α}}\,.
}
Explicitly, the $U(1)^2$ charges of the spinorial derivatives are
\equn{\begin{tabular}{c|cccccc}
$U(1)^2$ & $D^1_α$ & $D^2_α$ & $D^3_α$ & $\bar{D}_{1\dot{α}}$ & $\bar{D}_{2\dot{α}}$ & $\bar{D}_{3\dot{α}}$ \\
\hline
$q_1$ & $1$ & $-1$ & $0$ & $-1$ & $1$ & $0$ \\
$q_2$ & $1$ & $1$ & $-2$ & $-1$ & $-1$ & $2$
\end{tabular}
}

The harmonic superspace is an analytic subspace of the full superspace, where the coordinates $θ_1^α$ and $\bar{θ}^{2\dot{α}}$ do not appear explicitly in a given harmonic superfield $Φ^{(q_1,q_2)}(x,θ,u)$, \ie,
\equ{D^1_αΦ^{(q_1,q_2)}=\bar{D}_{2\dot{α}}Φ^{(q_1,q_2)}=0\,.
}
Note that these analytic constraints are preserved by the three harmonic derivatives $D^1_2,D^1_3,D^3_2$.

Finally, we can define an analytic measure $∫du\,dζ$ on harmonic superspace via the full superspace as follows:
\equ{∫d^4x\,d^{12}θ\,du ≡ ∫du\,dζ^{22}_{11}(D^1)^2(\bar{D}_2)^2 \quad ⇒ \quad ∫dζ^{22}_{11} = ∫d^4x_A\,(D^2)^2(D^3)^2(\bar{D}_1)^2(\bar{D}_3)^2\,.
}
We frequently use the notation $[D^4_ϑ]^{11}_{22}≡(D^1)^2(\bar{D}_2)^2$ to denote the four $θ$'s that are not part of the harmonic superspace. The $[D^4_ϑ]^{11}_{22}$ has $U(1)^2$ charge $(4,0)$, the negative of that for the measure $dζ^{22}_{11}$.

\subsection{SYM Action}
We do not review here the procedure for finding prepotentials of the $\N=3$ SYM in harmonic superspace but instead simply state the results. The $\N=3$ prepotentials are the gauge connections of the analyticity-preserving harmonic derivatives, \ie, we have three connections defined by $∇=D+\i A$. The gauge transformations read as usual: $δA=-∇λ$. The field strengths are introduced via the ``flat'' commutation relations as follows
\equ{\left[∇^1_2,∇^1_3\right] = F^{11}_{23}\,,\qquad \left[∇^3_2,∇^1_2\right] = F^{31}_{22}\,,\qquad \left[∇^1_3,∇^3_2\right] = ∇^1_2 + F^{1}_{2}\,.
}
The equations of motion are, of course, all $F=0$. These are generated simply by the following Chern-Simons-like action:
\equ{\S=\tr∫du\,dζ^{22}_{11}\,\(A^1_2F^1_2 +A^3_2F^{11}_{23} +A^1_3F^{31}_{22} -\i A^1_2[A^1_3,A^3_2]\).
\label{N3SYMact}}
Also, notice that one of the three prepotentials is related algebraically to the other two on-shell,
\equ{D^1_3A^3_2 -D^3_2A^1_3 +\i [A^1_3,A^3_2] =A^1_2\,,
\label{A12alg}}
from which we start the quantization procedure in the next section.

\section{\texorpdfstring{Quantizing SYM in ``Fermi-Feynman'' Gauge}{Quantizing SYM in "Fermi-Feynman" Gauge}}\label{sec:FFgauge}
The $\N=3$ SYM action \eqref{N3SYMact}, after substituting the algebraic equation defining $A^1_2$ \eqref{A12alg}, depends only on two harmonic connections and reads
\eqst{\S=\tr∫du\,dζ^{22}_{11}\,\Big\{(D^1_3A^3_2)^2 +(D^3_2A^1_3)^2 +2A^1_3(D^3_2D^1_3A^3_2) -2A^1_3D^1_2A^3_2 \\
+2\i[A^1_3,A^3_2]\(D^1_3A^3_2-D^3_2A^1_3\)-[A^1_3,A^3_2]^2\Big\}\,.
}
We choose the following gauge-fixing function:
\equ{\S_{gf}=-\tr∫du\,dζ^{22}_{11}\(D^1_3A^3_2+D^3_2A^1_3\)^2 =-\tr∫du\,dζ^{22}_{11}\left\{(D^1_3A^3_2)^2+(D^3_2A^1_3)^2 -2A^1_3(D^3_2D^1_3A^3_2)\right\},
\label{gffunexp}}
such that the gauge-fixed action for SYM in ``Fermi-Feynman'' gauge becomes
\equ{\S+\S_{gf}=\tr∫du\,dζ^{22}_{11}\,\left\{2A^1_3\(D^1_3D^3_2 +D^3_2D^1_3 -2D^1_2\)A^3_2 + 2\i[A^1_3,A^3_2]\(D^1_3A^3_2 - D^3_2A^1_3\)-[A^1_3,A^3_2]^2\right\},
\label{Avert}
}
where we used $[D^1_3,D^3_2]=D^1_2$ once. The ghost action follows from the BRST formalism straightforwardly by using $δA=-Dλ -\i[A,λ]$:
\equ{\S_{gh}= -\tr∫du\,dζ^{22}_{11}\,\left\{b^1_2\(D^1_3D^3_2+D^3_2D^1_3\)c +\i\bigl(D^1_3b^1_2\,[A^3_2,c] +D^3_2b^1_2\,[A^1_3,c]\bigr)\right\}.
\label{bcvert}
}

Having just introduced new fields, let us recap the $U(1)^2$ charges of all the fields here (which are straightforwardly deduced from the covariant derivatives):
\equn{\begin{tabular}{c|cccccc}
$U(1)^2$ & $A^1_3$ & $A^3_2$ & $A^1_2$ & $λ$ & $c$ & $b^1_2$ \\
\hline
$q_1$ & $1$ & $1$ & $2$ & $0$ & $0$ & $2$ \\
$q_2$ & $3$ & $-3$ & $0$ & $0$ & $0$ & $0$
\end{tabular}
}

\subsection{Propagators}
From the gauge-fixed SYM and ghost actions given above, we can derive the equations to solve for the Green's functions for vector and ghost superfields:
\eqss{(K^1_2)_0G^{0|1}_{0|2}(1,2) &=[δ_A]^{1|1}_{2|2}(1,2)\,, \\
(K^1_2)_0G^{1|0}_{2|0}(1,2) &=[δ_A]^{11|0}_{22|0}(1,2)\,, \\
(K^1_2)_{+1}G^{1|3}_{3|2}(1,2) &=[δ_A]^{11|3}_{23|2}(1,2)\,, \\
(K^1_2)_{-1}G^{3|1}_{2|3}(1,2) &=[δ_A]^{13|1}_{22|3}(1,2)\,,
\label{Greeneqs}}
where $(K^1_2)_a=(1+|a|)\(\{D^1_3,D^3_2\}+2 a D^1_2\)$ and the analytic delta functions explicitly read
\eqss{[δ_A]^{1|1}_{2|2}(1,2) &= δ(x_{12})[D^4_{vϑ}]^{11}_{22}δ^{12}(θ_{12})\,δ^{1|2}_{2|1}(u,v)=δ(x_{12})[D^4_{vϑ}]^{11}_{22}δ^{12}(θ_{12})\,\big({U^1} _1{\bar{U}_2}^{~2}\big)δ^6(u,v)\,, \\
[δ_A]^{11|0}_{22|0}(1,2) &= δ(x_{12})[D^4_{vϑ}]^{11}_{22}δ^{12}(θ_{12})\,δ^{11|22}_{22|11}(u,v)=δ(x_{12})[D^4_{vϑ}]^{11}_{22}δ^{12}(θ_{12})\,\big({U^1} _1{\bar{U}_2}^{~2}\big)^2δ^6(u,v)\,, \\
[δ_A]^{11|3}_{23|2}(1,2) &=δ(x_{12})[D^4_{vϑ}]^{11}_{22}δ^{12}(θ_{12})\,δ^{11|23}_{23|11}(u,v)=δ(x_{12})[D^4_{vϑ}]^{11}_{22}δ^{12}(θ_{12})\({U^1} _1\)^3 δ^6(u,v)\,, \\
[δ_A]^{13|1}_{22|3}(1,2) &=δ(x_{12})[D^4_{vϑ}]^{11}_{22}δ^{12}(θ_{12})\,δ^{13|22}_{22|13}(u,v)=δ(x_{12})[D^4_{vϑ}]^{11}_{22}δ^{12}(θ_{12})\({\bar{U}_2}^{~2}\)^3δ^6(u,v)\,.
\label{AnaDel}}
The general form of $G$'s which satisfy the Green's function equations then looks like
\eqs{\bk{c(1)b^1_2(2)} ≡ G^{0|1}_{0|2}(1,2) &=\F^{22|2}_{11|1}(u,v)\tfrac{1}{\square}[D^4_{uϑ}]^{11}_{22}[D^4_{vϑ}]^{11}_{22}δ^{12}(θ_{12})δ(x_{12})\,, \label{ghprop} \\
\bk{b^1_2(1)c(2)} ≡ G^{1|0}_{2|0}(1,2) &=\F^{2|22}_{1|11}(u,v)\tfrac{1}{\square}[D^4_{uϑ}]^{11}_{22}[D^4_{vϑ}]^{11}_{22}δ^{12}(θ_{12})δ(x_{12})\,, \label{ghpropf} \\
\bk{A^1_3(1)A^3_2(2)} ≡ G^{1|3}_{3|2}(1,2) &=\F^{22|23}_{13|11}(u,v)\tfrac{1}{\square}[D^4_{uϑ}]^{11}_{22}[D^4_{vϑ}]^{11}_{22}δ^{12}(θ_{12})δ(x_{12})\,, \label{vecprop} \\
\bk{A^3_2(1)A^1_3(2)} ≡ G^{3|1}_{2|3}(1,2) &=\F^{23|22}_{11|13}(u,v)\tfrac{1}{\square}[D^4_{uϑ}]^{11}_{22}[D^4_{vϑ}]^{11}_{22}δ^{12}(θ_{12})δ(x_{12})\,, \label{vecpropF}
}
such that
\equ{(K^1_2)_a\F^{·|·}_{·|·}(u,v) =\half (D^2_1)^2δ^{·|·}_{·|·}(u,v)\,.
\label{genexp}}
The $δ^{·|·}_{·|·}(u,v)$ functions are the same $δ$-functions appearing in the corresponding $[δ_A]^{·|·}_{·|·}(1,2)$ defined in \eqref{AnaDel}. Equation \eqref{genexp} is motivated by the identity\footnote{We will suppress the $SU(3)$ ``indices'' on $[D^4_{ϑ}]^{11}_{22}$ from now on.}
\equ{(D^2_1)^2D^4_{uϑ}D^4_{vϑ}δ^6(u,v) =2\square D^4_{vϑ}δ^6(u,v)\,,
\label{D2D4D4box}}
which can be used to prove that \eqref{ghprop}--\eqref{vecpropF} indeed satisfy \eqref{Greeneqs}.

In order to make the above equations simpler and more tractable, we choose the following ``independent'' internal coordinates in the $\{\Uv{a}{b},\Ub{a}{b}\}$ basis:
\equ{U^1_{~1}, U^1_{~3}, U^2_{~1}, U^3_{~1}, \bar{U}_2^{~1}, \bar{U}_2^{~2}, \bar{U}_2^{~3}, \bar{U}_3^{~2}\,,
}
and the `zero charge' $\delta$-function in these coordinates reads
\equ{\delta^6(u,v) = πU^1_{~1} \bar{U}_2^{~2} \delta (U^1_{~3}) \delta (U^3_{~1}) \delta (\bar{U}_2^{~3}) \delta (\bar{U}_3^{~2}) \delta (U^2_{~1}) \delta (\bar{U}_2^{~1})\,.
\label{delta^6}}
The rest of the $U$-coordinates can be written in terms of the chosen ones as follows:
\equ{\begin{aligned}
\Uv{1}{2} &= -\frac{\Uv{1}{1}\Ub{2}{1} + \Uv{1}{3}\Ub{2}{3}}{\Ub{2}{2}}\,,\qquad && \Uv{2}{2} = - \frac{\Uv{1}{1}\Uv{2}{1}\Ub{2}{1} + \Uv{1}{3}\Uv{2}{1}\Ub{2}{3} - \Ub{2}{3}\Ub{3}{2} - \Uv{1}{1}}{\Uv{1}{1}\Ub{2}{2}}\,, \\
\Uv{2}{3} &= \frac{\Uv{1}{3}\Uv{2}{1} - \Ub{3}{2}}{\Uv{1}{1}}\,,\qquad && \Uv{3}{2} = - \frac{\Uv{1}{1}\Uv{3}{1}\Ub{2}{1} + \Uv{1}{3}\Uv{3}{1}\Ub{2}{3} + \Ub{2}{2}\Ub{2}{3}}{\Uv{1}{1}\Ub{2}{2}}\,, \\
\Uv{3}{3} &= \frac{\Uv{1}{3}\Uv{3}{1} + \Ub{2}{2}}{\Uv{1}{1}}\,,\qquad && \Ub{1}{1} = \frac{-\Uv{2}{1}\Ub{2}{1}\Ub{2}{2} - \Uv{3}{1}\Ub{2}{1}\Ub{3}{2} + \Uv{1}{3}\Uv{3}{1} + \Ub{2}{2}}{\Uv{1}{1}\Ub{2}{2}}\,, \\
\Ub{1}{2} &= -\frac{\Uv{2}{1}\Ub{2}{2} + \Uv{3}{1}\Ub{3}{2}}{\Uv{1}{1}}\,,\qquad && \Ub{1}{3} = - \frac{\Uv{2}{1}\Ub{2}{2}\Ub{2}{3} + \Uv{3}{1}\Ub{2}{3}\Ub{3}{2} + \Uv{1}{1}\Uv{3}{1}}{\Uv{1}{1}\Ub{2}{2}}\,, \\
\Ub{3}{1} &= \frac{\Ub{2}{1}\Ub{3}{2} - \Uv{1}{3}}{\Ub{2}{2}}\,,\qquad && \Ub{3}{3} = \frac{\Ub{2}{3}\Ub{3}{2} + \Uv{1}{1}}{\Ub{2}{2}}\,·
\end{aligned}}
The differential operators get modified as well, leading to the following expressions:
\begin{align}
D^1_{~2} &= U^1_{~1}\frac{\partial}{\partial U^2_{~1}}\,,\qquad D^1_{~3} = U^1_{~1}\frac{\partial}{\partial U^3_{~1}}\,,\qquad D^3_{~2} = U^3_{~1}\frac{\partial}{\partial U^2_{~1}} - \bar{U}_2^{~2}\frac{\partial}{\partial \bar{U}_3^{~2}}\,,\\
D^2_{~1} &= U^2_{~1}\frac{\partial}{\partial U^1_{~1}} + U^2_{~3}\frac{\partial}{\partial U^1_{~3}} - \bar{U}_1^{~1}\frac{\partial}{\partial \bar{U}_2^{~1}}- \bar{U}_1^{~2}\frac{\partial}{\partial \bar{U}_2^{~2}} - \bar{U}_1^{~3}\frac{\partial}{\partial \bar{U}_2^{~3}} \nn
&= U^2_{~1}\frac{\partial}{\partial U^1_{~1}} + \left(\frac{U^1_{~3}U^2_{~1} - \bar{U}_3^{~2}}{U^1_{~1}}\right)\frac{\partial}{\partial U^1_{~3}} -\left(\frac{-U^2_{~1}\bar{U}_2^{~1}\bar{U}_2^{~2} - U^3_{~1}\bar{U}_2^{~1}\bar{U}_3^{~2} + U^1_{~3}U^3_{~1}+\bar{U}_2^{~2}}{U^1_{~1}\bar{U}_2^{~2}}\right)\frac{\partial}{\partial \bar{U}_2^{~1}} \nn
&\quad +\left(\frac{U^2_{~1}\bar{U}_2^{~2} + U^3_{~1}\bar{U}_3^{~2}}{U^1_{~1}}\right)\frac{\partial}{\partial \bar{U}_2^{~2}} + \left(\frac{U^2_{~1}\bar{U}_2^{~2}\bar{U}_2^{~3} + U^3_{~1}\bar{U}_2^{~3}\bar{U}_3^{~2} + U^1_{~1}U^3_{~1}}{U^1_{~1}\bar{U}_2^{~2}}\right)\frac{\partial}{\partial \bar{U}_2^{~3}}\,, \\
D^3_{~1} &= U^3_{~1}\frac{\partial}{\partial U^1_{~1}} + U^3_{~3}\frac{\partial}{\partial U^1_{~3}} - \bar{U}_1^{~2}\frac{\partial}{\partial \bar{U}_3^{~2}} \nn
&= U^3_{~1}\frac{\partial}{\partial U^1_{~1}} + \left(\frac{U^1_{~3}U^3_{~1} + \bar{U}_2^{~2}}{U^1_{~1}}\right)\frac{\partial}{\partial U^1_{~3}} + \left(\frac{U^2_{~1}\bar{U}_2^{~2} + U^3_{~1}\bar{U}_3^{~2}}{U^1_{~1}}\right)\frac{\partial}{\partial \bar{U}_3^{~2}}\,, \\
D^2_{~3}&=U^2_{~1}\frac{\partial}{\partial U^3_{~1}} - \bar{U}_3^{~1}\frac{\partial}{\partial \bar{U}_2^{~1}} - \bar{U}_3^{~2}\frac{\partial}{\partial \bar{U}_2^{~2}} - \bar{U}_3^{~3}\frac{\partial}{\partial \bar{U}_2^{~3}} \nn
&= U^2_{~1}\frac{\partial}{\partial U^3_{~1}} + \left(\frac{U^1_{~3} - \bar{U}_2^{~1}\bar{U}_3^{~2} }{\bar{U}_2^{~2}}\right)\frac{\partial}{\partial \bar{U}_2^{~1}} - \bar{U}_3^{~2}\frac{\partial}{\partial \bar{U}_2^{~2}}- \left(\frac{U^1_{~1} + \bar{U}_2^{~3}\bar{U}_3^{~2} }{\bar{U}_2^{~2}}\right)\frac{\partial}{\partial \bar{U}_2^{~3}}\,·
\end{align}
Using all these expressions, we can write \eqref{genexp} explicitly in the following form
\equ{(1+|a|)\left[U^3{}_1\frac{∂^2}{∂U^2{}_1∂U^3{}_1} -\bar{U}_2{}^2\frac{∂^2}{∂\bar{U}_3{}^2∂U^3{}_1}+\(a+\frac{1}{2}\)\frac{∂}{∂U^2{}_1}\right]\F^{·|·}_{·|·}=Δ^{·|·}_{·|·}\delta(U^3{}_1) \delta(U^2{}_1) \delta(\bar{U}_3{}^2)\,,
\label{Umaineq}}
where
\equn{Δ^{·|·}_{·|·} = \tfrac{π}{2}\(U^1{}_1\)^{n_1}\(\bar{U}_2{}^2\)^{n_2} \delta(U^1{}_3) \delta(\bar{U}_2{}^3)   \delta''(\bar{U}_2{}^1) \quad\text{with } (n_1,n_2)=\left\{\begin{array}{cl} (-1,2) & \text{for } \bk{c(1)b^1_2(2)} \\
(0,3) & \text{for } \bk{b^1_2(1)c(2)} \\
(1,1) & \text{for } \bk{A^1_3(1)A^3_2(2)} \\
(-2,4) & \text{for } \bk{A^3_2(1)A^1_3(2)}\,.
\end{array}\right.
}
Relabelling $U^3{}_1=x$, $U^1{}_3=\hat{x}$, $U^2{}_1=y$, $\bar{U}_2{}^1=\bar{y}$, $\bar{U}_3{}^2=z$, $\bar{U}_2{}^3=\hat{z}$, $\bar{U}_2{}^2=A$, $U^1{}_1=B$ and $a+\frac{1}{2}=b$, we get a simple looking partial differential equation
\equ{\left(x ∂_x ∂_y - A ∂_x ∂_z + b ∂_y \right) \F^{·|·}_{·|·} = \tfrac{π}{2(1+|a|)} B^{n_1}A^{n_2}\delta(x)δ(\hat{x})\delta(y)δ''(\bar{y})\delta(z)δ(\hat{z})\,.
\label{Smaineq}
}
We solve it by choosing an Ansatz of the form
\equ{\F^{·|·}_{·|·} =C_a A^{p}B^{q}\frac{y^r\bar{y}^s}{\big(\bar{y}+\frac{ε^2}{y}\big)^3}\(\frac{Ay+xz}{B}\)^t\(\frac{B\bar{y}+\hat{x}\hat{z}}{A}\)^{-1},
\label{soltn}
}
where $ε$ is an infinitesimal parameter and the exponents $p,q,r,s,t$ along with the normalization factor $C_a$ are to be determined. Plugging \eqref{soltn} into the LHS of \eqref{Smaineq}, we find that the values $t=-b$, $r=b-1$, $s=0$ simplify the expression to a single term as follows:
\eqs{\text{LHS of \eqref{Smaineq}} &=\frac{3C_a b A^{p+2}B^{q+b}y^{b+2}(A y+x z)^{-b}ε^2}{(A y+x z)(B\bar{y}+\hat{x}\hat{z})(y\bar{y}+ε^2)^4} \nn
&= \frac{C_a b A^{p+2}B^{q+b}\big(A +\frac{x z}{y}\big)^{-b}}{2(A y+x z)(B\bar{y}+\hat{x}\hat{z})}πδ(y)δ''(\bar{y}) \nn
&=\tfrac{π}{2}C_a b A^{p+2}B^{q+b}\big(A +\tfrac{x z}{y}\big)^{-b}\delta(x)\delta(z)δ(\hat{x})δ(\hat{z})\delta(y)δ''(\bar{y}) \nn
&=\tfrac{π}{2}C_a b B^{q+b}A^{p+2-b}\delta(x)δ(\hat{x})\delta(y)δ''(\bar{y})\delta(z)δ(\hat{z})\,,
\label{solcheck}}
where we use the fact that $δ$-functions for $y,\bar{y}$ are produced in the limit of $ε→0$ and $y→0$ via the following identity:
\eqs{\frac{(m+1)!(-y)^m ε^2}{(y\bar{y}+ε^2)^{m+2}}→πδ(y)δ^{(m)}(\bar{y})\,,
\label{deltaID}}
and $δ(z)=\frac{1}{z}$, etc., for rest of the four nonconjugate complex variables. Now, comparing \eqref{solcheck} to the RHS of \eqref{Smaineq}, we can deduce that $p=n_2+b-2$, $q=n_1-b$ and $C_a=\frac{1}{b(1+|a|)}\,·$ Thus, the final form of $\F^{·|·}_{·|·}$ that solves \eqref{Smaineq} reads
\equ{\F^{·|·}_{·|·} =\frac{1}{b(1+|a|)} A^{n_2+b-2}B^{n_1-b}\frac{y^{b-1}}{\big(\bar{y}+\frac{ε^2}{y}\big)^3}\(\frac{Ay+xz}{B}\)^{-b}\(\frac{B\bar{y}+\hat{x}\hat{z}}{A}\)^{-1}.
}
Finally, the complete propagators in terms of the $U$-variables read as follows:
\equ{G^{·|·}_{·|·}(1,2)=\frac{(\Ub{2}{2})^{n_2+b-2}(\Uv{1}{1})^{n_1-b}}{b(1+|a|)}\frac{(U^2{}_1)^{b-1}}{\big(\bar{U}_2{}^1+\frac{ε^2}{U^2{}_1}\big)^3}\frac{\(-\bar{U}_1{}^2\)^{-b}}{\(-U^1{}_2\)}\frac{1}{\square}D^4_{uϑ}D^4_{vϑ}δ^{12}(θ_{12})δ(x_{12})\,.
\label{FinProp}
}

\subsection{Feynman Rules}
The Feynman rules are now derived as usual. The vector and ghost propagators are given in \eqref{FinProp} but we reproduce them here individually with explicit harmonic factors in momentum space (replace $\frac{1}{\square}δ(x_{12}) → \frac{1}{-k^2}$):
\eqs{\bk{c(1)b^1_2(2)} &= \frac{2(\Ub{2}{2})^{\frac{1}{2}}}{(U^1{}_1)^{\frac{3}{2}}}\frac{(U^2{}_1)^{-\frac{1}{2}}}{\big(\bar{U}_2{}^1+\frac{ε^2}{U^2{}_1}\big)^3}\frac{\(-\bar{U}_1{}^2\)^{-\frac{1}{2}}}{U^1{}_2}\frac{1}{k^2}D^4_{uϑ}D^4_{vϑ}δ^{12}(θ_{12})\,, \\
\bk{b^1_2(1)c(2)} &= \frac{2(\Ub{2}{2})^{\frac{3}{2}}}{(\Uv{1}{1})^{\frac{1}{2}}}\frac{(U^2{}_1)^{-\frac{1}{2}}}{\big(\bar{U}_2{}^1+\frac{ε^2}{U^2{}_1}\big)^3}\frac{\(-\bar{U}_1{}^2\)^{-\frac{1}{2}}}{U^1{}_2}\frac{1}{k^2}D^4_{uϑ}D^4_{vϑ}δ^{12}(θ_{12})\,, \\
\bk{A^1_3(1)A^3_2(2)} &= \frac{(\Ub{2}{2})^{\frac{1}{2}}}{3(\Uv{1}{1})^{\frac{1}{2}}}\frac{(U^2{}_1)^{\frac{1}{2}}}{\big(\bar{U}_2{}^1+\frac{ε^2}{U^2{}_1}\big)^3}\frac{\(-\bar{U}_1{}^2\)^{-\frac{3}{2}}}{U^1{}_2}\frac{1}{k^2}D^4_{uϑ}D^4_{vϑ}δ^{12}(θ_{12})\,, \\
\bk{A^3_2(1)A^1_3(2)} &= -\frac{(\Ub{2}{2})^{\frac{3}{2}}}{(\Uv{1}{1})^{\frac{3}{2}}}\frac{(U^2{}_1)^{-\frac{3}{2}}}{\big(\bar{U}_2{}^1+\frac{ε^2}{U^2{}_1}\big)^3}\frac{\(-\bar{U}_1{}^2\)^{\frac{1}{2}}}{U^1{}_2}\frac{1}{k^2}D^4_{uϑ}D^4_{vϑ}δ^{12}(θ_{12})\,,
}
where we have kept the $ε$-prescription explicit. The vertices can be read from \eqref{Avert} and \eqref{bcvert}:
\eqsg{\bk{(A^1_3)^a(A^3_2)^b(A^3_2)^c} \,/\, \bk{(A^3_2)^a(A^1_3)^b(A^1_3)^c} &\quad→\quad 2\tint du\,d^8θ\,f^{abc}\left[D^1_3 \,/\, D^3_2\right] \\
\bk{(A^1_3)^a(A^3_2)^b(A^1_3)^c(A^3_2)^d} &\quad→\quad \i\tint du\,d^8θ\,f^{abe}f^{ecd} \\
\bk{(A^3_2)^ac^b(b^1_2)^c} \,/\, \bk{(A^1_3)^ac^b(b^1_2)^c} &\quad→\quad -\tint du\,d^8θ\,f^{abc}\left[D^1_3 \,/\, D^3_2\right],
\label{Qvert}}
where $∫d^8θ≡(D^2)^2(D^3)^2(\bar{D}_1)^2(\bar{D}_3)^2$, $f^{abc}$ are structure constants of the gauge group and the harmonic derivatives act on the leg corresponding to the group index `$c$'.

\paragraph{Evaluating loop graphs.} Let us first focus on 1-loop graphs. We can generate $∫d^{12}θ$ from the analytic measure at vertices by taking off one factor of $D^4_ϑ$ from the propagators. After this, we are left with $(v_3+v_4)$ $∫d^{12}θ$ integrals from 3- and 4-point vertices, $p$ $D^4_{ϑ}δ^{12}(θ)$'s from propagators. As usual, we need to saturate all but one $∫d^{12}θ$, which means we need to kill one of the $δ^{12}(θ)$. This is achieved by using three $D^4_{ϑ}$'s and the identity
\vspace*{-5mm}
\begingroup
\eqst{\hspace*{-3mm}D^4_{wϑ}D^4_{vϑ}D^4_{uϑ} \\
=\left[\((W^1{}_3)^2(\bar{W}_2{}^3)^2(V^1{}_2)^2(\bar{V}_2{}^1)^2 +(W^1{}_3)^2(\bar{W}_2{}^1)^2(V^1{}_2)^2(\bar{V}_2{}^3)^2 +(W^1{}_2)^2(\bar{W}_2{}^3)^2(V^1{}_3)^2(\bar{V}_2{}^1)^2 \right.\right. \\
\left.\left.+(W^1{}_2)^2(\bar{W}_2{}^1)^2(V^1{}_3)^2(\bar{V}_2{}^3)^2\)D^{8}_{uθ} -\i (W^1{}_2)^2(\bar{W}_2{}^3)^2(V^1{}_3)^2(\bar{V}_2{}^1)^2 D^3·∂·\bar{D}_3 (D^2)^2(\bar{D}_1)^2 \right. \\
\left.-\half \square\big((W^1{}_2)^2(\bar{W}_2{}^3)^2(V^1{}_3)^2(\bar{V}_2{}^1)^2(D^2)^2(\bar{D}_1)^2 +(W^1{}_3)^2(\bar{W}_2{}^3)^2(V^1{}_3)^2(\bar{V}_2{}^1)^2(D^3)^2(\bar{D}_1)^2 \right. \\
+(W^1{}_2)^2(\bar{W}_2{}^3)^2(V^1{}_3)^2(\bar{V}_2{}^3)^2(D^2)^2(\bar{D}_3)^2 +(W^1{}_3)^2(\bar{W}_2{}^3)^2(V^1{}_3)^2(\bar{V}_2{}^3)^2(D^3)^2(\bar{D}_3)^2\big)\big]D^4_{uϑ}\,,
\label{3dth4s}}
\endgroup

\vspace*{-3mm}
\noindent where $D^8_θ≡(D^2)^2(D^3)^2(\bar{D}_1)^2(\bar{D}_3)^2$, $V^a{}_b≡v^a\bar{u}_b$, $\bar{W}_a{}^b≡\bar{w}_au^b$, etc. The first term which contains $D^8_{θ}D^4_{ϑ}$ [recall that $D^4_{ϑ}≡(D^1)^2(\bar{D}_2)^2$] can be used to kill one $δ^{12}(θ)$. This means that a 2-point function trivially vanishes as it does not have enough $D^4_{ϑ}$'s, and a 3-point function cannot have any divergent piece due to the presence of three $\square$'s in the denominator, \ie, $∫\frac{d^4k}{(k^2)^3}$ is finite. In fact, no higher-point function can have any divergent piece at 1-loop because the numerator can generate at most $(p-3)$ $\square$'s (together with a $D^8_θ$) compared to $p$ $\square$'s in the denominator, so the difference is always $n≥3$, \ie, $∫\frac{d^4k}{(k^2)^n}$ is finite.

This power-counting readily generalizes to multiloop graphs because each 1-loop subgraph needs to follow this procedure of $D$ algebra, and hence the whole graph is rendered finite. This proves the nonrenormalization theorem at all loops for $\N=3$ SYM or, equivalently, $\N=4$ SYM.

\section{Quantizing SYM in Background Field Gauge}\label{sec:BFgauge}
The computation of finite terms for loop graphs is still cumbersome with the Feynman rules discussed in the previous section because manifestly covariant expressions are not obtained for individual graphs. For that purpose, we develop the background field formalism in this section.

Let us gauge covariantize all the differential operators $(D→∇=D+\i A)$. Then we do a background splitting of the connections in a straightforward manner: $A→\A_{bg}+\a_{q}$, where the subscripts will be suppressed in favour of self-explanatory fonts. Next, we choose different representations for these connections: the ``real'' rep for background $\A$'s, meaning that the three harmonic connections vanish $\(\A^1_2=\A^1_3=\A^3_2=0\)$, and the ``analytic'' rep for quantum $\a$'s, meaning that the four fermionic connections vanish $\(\a^1_α=\bar{\a}_{2\dot{α}}=0 ⇒ D_ϑ→∇_ϑ≡\D_ϑ\)$.\footnote{Such a choice was used in the case of $\N=2$ projective superspace to construct the background field formalism \cite{Jain:2013hua}. It ensures that the effective action is independent of background fields with dimension $0$ (like the harmonic connections), which is required for the nonrenormalization theorems to hold \cite{Grisaru:1982zh}. However, the simple splitting of $\N=3$ prepotentials is reminiscent of the background field formalism for the $\N=2$ harmonic superspace developed in \cite{Buchbinder:1997ya}, with further refinements and explicit calculations appearing in \cite{Buchbinder:1997ib,Buchbinder:1998np}.} Let us write down the consequences of these choices on connections and field strengths from various commutators:
\eqsg{\{∇^a_α,\bar{∇}_{b\dot{β}}\} &=\i δ^a_b∇_{α\dot{β}} \qquad \text{(``unchanged'')} \\
\{∇^a_α,∇^b_β\} &= ε_{αβ}\bar{W}^{ab} \\
\{\bar{∇}_{a\dot{α}},\bar{∇}_{b\dot{β}}\} &= ε_{\dot{α}\dot{β}}W_{ab} \\
[∇^a_b,∇^c_α] &=δ^c_b∇^a_α \\
[∇^a_b,\bar{∇}_{c\dot{α}}] &=-δ^a_c\bar{∇}_{b\dot{α}} \\
[∇^a_b,∇^c_d] &= \delta^c_b∇^a_d - \delta^a_d∇^c_b\,.
\label{backcomm}}
Of course, the $W$'s are antisymmetric in the two indices and they satisfy a few Bianchi identities along with some analytic plus harmonic constraints \cite{Ivanov:2001ec,Buchbinder:2016xeq}. The most relevant identity for us is
\equ{D^a_αW_{bc}=\half(δ^a_b D^k_αW_{kc} -δ^a_c D^k_αW_{kb})\,.
\label{BianchId}}
Another thing to note is that the spinorial background-covariant derivatives $\D^1_α$ and $\bar{\D}_{2\dot{α}}$ still possess the structure of \eqref{intgstr}, so that the ``background analytic'' superfields can be defined: $\D^1_αΦ=\bar{\D}_{2\dot{α}}Φ=0$. Moreover, the fourth and fifth equations of \eqref{backcomm} tell us that the harmonic connections we are most interested in are now background analytic. We will also need the following identity defining a generalized d'Alembertian:
\equ{\D^4_ϑ (∇^2_1)^2\D^4_ϑ =2\,\wh{\square}\,\D^4_ϑ = 2\left[(\square -2\bar{W}^{12}W_{21}) -\bar{∇}_1·\bar{∇}_a\bar{W}^{1a} +2∇^2·\D^1W_{21} +∇^2_1(\D^1)^2W_{21}\right]\D^4_ϑ\,.
\label{gendal}}

\subsection{Feynman Rules}
Since the harmonic connections appearing in the action as well as the ghosts are purely quantum superfields but background analytic, only $\square$ changes to $\wh{\square}$ in the propagators derived in the previous section, whereas the vertices in the Lagrangian remain the same. However, we can expand the $\wh{\square}$ to get vertices with explicit field strengths, which leads to covariant results in the loop calculations directly. With this structure, we can write down the background Feynman rules as follows:
\eqsn{\text{Propagators: }& \left\{\begin{array}{cl} \F^{·|·}_{·|·}(u,v)\frac{1}{-k^2}\D^4_{uϑ}\D^4_{vϑ}δ^{12}(θ_{12}) & \text{ignore background vertices} \\
\frac{1}{-k^2}δ^8(θ_{12})δ^6(u,v) & \text{consider all such vertices but one,}
\end{array}\right. \\
\text{All but one background vertices: }& \tint du\,d^8θ\(\wh{\square}-\square_0\), \\
\text{One background vertex: }& \tint du\,dv\,d^8θ\(K^1_2\)_aδ^6(u,v)\,, \\
\text{All quantum vertices: }& \text{Same as \eqref{Qvert}.}
}
Note that the background d'Alembertian is expanded as $\square≡\D^{α\dot{α}}\D_{α\dot{α}}=\square_0+⋯$, where $\square_0$ corresponds to $-k^2$ in the momentum space.

\paragraph{1-loop graph computation.} Let us focus on the computation of a 4-point function here. It is finite and, from symmetry arguments of Section 5.5 of \cite{Buchbinder:2016xeq}, it is known to look like $(\bar{W}^{13}W_{23})^2$ in $\N=3$ harmonic superspace.
\fig{h!}{\includegraphics[scale=1]{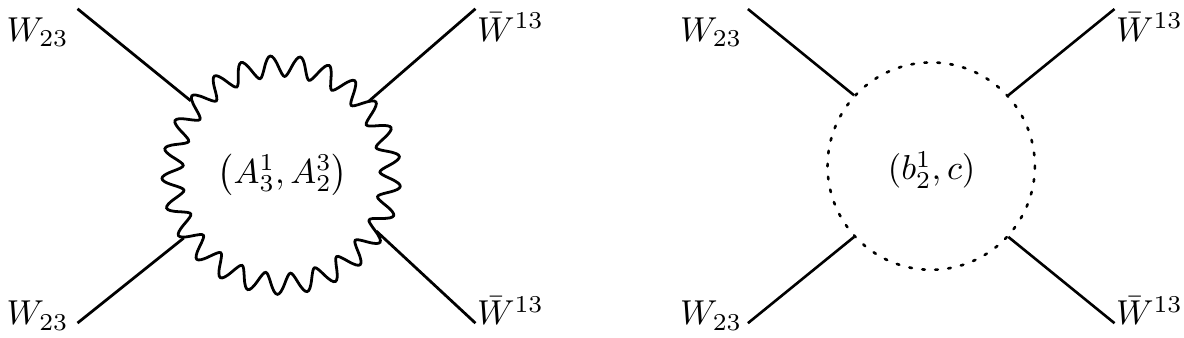}
\caption{1-loop 4-point graph with external background field strengths.}
\label{fig:4p4v1L}}

We can obtain such a contribution by evaluating a bubble graph and expanding the $(\wh{\square}-\square_0)$-vertex factors to get the relevant $W$'s using \eqref{gendal} and \eqref{BianchId}. The above Feynman rules give the following expression for the graph with vector loop shown in Figure \ref{fig:4p4v1L} (after performing $θ$ and $u$ integrals at two of the vertices):
\eqst{Γ_4^{(A)} \sim \hat{\A}_4∫du\,dv∫d^8θ_{1,2}\(\bar{∇}_1·\bar{∇}_3\bar{W}^{13}(p_1)\)\(\bar{∇}_1·\bar{∇}_3\bar{W}^{13}(p_2)\)\((U^2{}_1U^1{}_2)∇^2·∇^3W_{23}(p_3)\) \\
×δ^8(θ_{12})(K^1_2)_{-1}\,\F^{22|23}_{13|11}(u,v)\,\D_{uϑ}^4(p₄)\D_{vϑ}^4(p₄)δ^{12}(θ_{12})δ^6(u,v) \,,
}
where $\hat{\A}_4\sim∫d^4k\frac{1}{(k_1^2)(k_2^2)(k_3^2)(k_4^2)}$ is the scalar box integral with the subscripts on loop momentum $k$ denoting the external momenta ($p_i$) dependence. Note that we had to partially integrate $(K^1_2)_{+1}$ so that it changed to $(K^1_2)_{-1}$ acting on $\F^{·|·}_{·|·}$, which gives, using \eqref{Umaineq}, $(K^1_2)_{-1}\,\F^{22|23}_{13|11}(u,v)=\frac{1}{2}(D^2_1)^2δ^6(u,v) -4\frac{∂}{∂U^2{}_1}\F^{22|23}_{13|11}$. The first term leads to a harmonic singularity with two $δ^6(u,v)$-functions, but this singularity will cancel with the analogous contribution $Γ_4^{(bc)}$ from the ghost loop in Figure \ref{fig:4p4v1L}. So we focus only on the second term, which has no analogue from the ghost loop graph (as $a=0$) and thus gives the complete four-point function. To get rid of the $δ^{12}(θ_{12})$-function, eight spinorial derivatives should be gathered off of $W$'s in addition to $\D^4_{ϑ}$ as follows\footnote{The requirement of collecting eight spinorial derivatives to get $∇^8_θ$ [which is analogous to extracting $D^8_θ$ from \eqref{3dth4s} while evaluating graphs in ``Fermi-Feynman'' gauge] is sufficient to make $Γ_2=Γ_3=0$ identically at 1-loop.}:
\eqs{Γ_4 &\sim\, \hat{\A}_4∫du\,dv∫d^8θ_{1,2}\,\bar{W}^{13}(p_1)\bar{W}^{13}(p_2)W_{23}(p_3)\,δ^8(θ_{12})\,D^1_2\F^{22|23}_{13|11}(u,v) \nn
&\qquad ×(\bar{∇}_1)^2(\bar{∇}_3)^2\((U^2{}_1U^1{}_2)∇^2·∇^3\)\(\bar{U}_2{}^1\bar{U}_2{}^1U^2{}_1U^1{}_2∇^2·∇^3W_{23}(p_4)\)\D_{vϑ}^4(p_4)δ^{12}(θ_{12})δ^6(u,v) \nn
&\sim\, \hat{\A}_4∫du∫d^8θ\,\bar{W}^{13}(p_1)\bar{W}^{13}(p_2)W_{23}(p_3)W_{23}(p_4)\left[\(\frac{U^1{}_1}{(\bar{U}_2{}^1)^4(U^2{}_1)^2}\)(U^2{}_1U^1{}_2)^2(\bar{U}_2{}^1)^2\right]_{|_{u→v}} \nn
&\sim\ \hat{\A}_4∫du∫d^8θ\,\bar{W}^{13}(p_1)\bar{W}^{13}(p_2)W_{23}(p_3)W_{23}(p_4)\left[\frac{(U^1{}_2)^2}{(\bar{U}_2{}^1)^2}\right]_{|_{u→v}},
\label{V4p}}
where we use $∇^8_θ\D^4_ϑδ^{12}(θ_{12})=1$ in the second step. In the last step, the apparent harmonic singularity cancels as we take the limit $u→v$, leading to the expected result for the 1-loop effective action,
\equ{Γ_4 \sim ∫du\,dζ^{22}_{11}\,\hat{\A}_4\(\bar{W}^{13}W_{23}\)^2.
}

\section{Discussion}\label{sec:discuss}
We have introduced a new gauge-fixing action to quantize $\N=3$ SYM in $\N=3$ harmonic superspace. This leads to simpler (and fewer) propagators for vector and ghost superfields in a ``Fermi-Feynman'' gauge. These are sufficient to prove the nonrenormalization theorem for $\N=3,4$ SYM at all loops. However, computation of loop graphs beyond the divergent terms can be simplified more with the background field formalism. With the background Feynman rules in hand, we have computed the 1-loop 4-point contribution to the effective action, which gets a finite contribution purely from the vector loop diagram.

The way that we derived the harmonic propagators here is reminiscent of how $\N=2$ projective superspace\cite{Lindstrom:1987ks,Lindstrom:1989ne} is derived from $\N=2$ harmonic superspace\cite{Galperin:1984av,Galperin:2007wpa} in \cite{Jain:2009aj}: by choosing a special parameterization of the harmonic R-symmetry coordinates on $SU(2)/U(1)≃S^2$ to obtain a single (complex) coordinate on $CP^1$ that forms the internal coordinate for the projective case.\footnote{The relation between these two hyperspaces was explored from different points of view in \cite{Kuzenko:1998xm,Butter:2012ta}.} Even though we did not take this route to its full conclusion, it should be possible to derive a $\N=3$ projective superspace in such a way that it simplifies the $∫du$-integrals to something more tractable. For example, viewing the R-symmetry coset $SU(3)/U(1)^2 ≃ [SU(3)/\(SU(2)×U(1)\)] × [SU(2)/U(1)] ≃ CP^2×CP^1$, one can expect reducing the six ($x,\hat{x},y,\bar{y},z,\hat{z}$) R-symmetry coordinates of $\N=3$ harmonic superspace to only three ($x,y,z$) for a possible $\N=3$ projective superspace. This should then be followed by a projection of the harmonic gauge condition and equations of motion for the gauge and ghost fields to the projective superspace, which we leave for future work.

\section*{\centering Acknowledgements}
DJ thanks Daniel Butter and Yu-tin Huang for the encouraging discussions. DJ would also like to thank the Simons Center for Geometry and Physics for its hospitality during the ``2016 Simons Summer Workshop'', where a part of this work was done. CYJ is supported in part by the National Center for Theoretical Sciences and the Ministry of Science and Technology, Taiwan, through MOST Grant No. 109-2811-M-005-509. WS is supported by NSF Grant No. PHY-1915093.

\references{refsN3HS}

\end{document}